\documentclass[preprint,showpacs,preprintnumbers,amsmath,amssymb]{revtex4}
\usepackage{epsfig}
\usepackage{graphicx}
\usepackage{dcolumn}
\usepackage{bm}
\usepackage{threeparttable}

\def\beq{\begin{equation}}
\def\eeq{\end{equation}}
\def\eeqn{\end{equation}}
\newcommand\iden{\leavevmode\hbox{\small1\normalsize\kern-.33em1}}

\newcommand{\bea} {\begin{eqnarray}}
\newcommand{\eea} {\end{eqnarray}}

\let\jnfont=\rm
\def\NPB#1,{{\jnfont Nucl.\ Phys.\ B }{\bf #1},}
\def\PLB#1,{{\jnfont Phys.\ Lett.\ B }{\bf #1},}
\def\EPJC#1,{{\jnfont Eur.\ Phys.\ Jour.\ C }{\bf #1},}
\def\PRD#1,{{\jnfont Phys.\ Rev.\ D }{\bf #1},}
\def\PRL#1,{{\jnfont Phys.\ Rev.\ Lett.\ }{\bf #1},}
\def\MPLA#1,{{\jnfont Mod.\ Phys.\ Lett.\ A }{\bf #1},}
\def\JPG#1,{{\jnfont J.\ Phys.\ G }{\bf #1},}
\def\CTP#1,{{\jnfont Commun.\ Theor.\ Phys.\ }{\bf #1},}
\def\JHEP#1,{{\jnfont JHEP \ }{\bf #1},}
\def\NPPS#1,{{\jnfont Nucl.\ Phys.\ Proc.\ Suppl.\ }{\bf #1},}
\def\CPC#1,{{\jnfont Comput.\ Phys.\ Commun.\ }{\bf #1},}
\def\CPL#1,{{\jnfont Chin.\ Phys.\ Lett. }{\bf #1},}
\def\APPB#1,{{\jnfont Acta\ Phys.\ Polon.\ B }{\bf #1},}

\def\lsim{\raise0.3ex\hbox{$<$\kern-0.75em\raise-1.1ex\hbox{$\sim$}}}
\def\gsim{\raise0.3ex\hbox{$>$\kern-0.75em\raise-1.1ex\hbox{$\sim$}}}
\def\PR#1,{{\jnfont Phys.\ Rept. }{\bf #1},}
\def\CHC#1,{{\jnfont Chin.\ Phys.\ C }{\bf #1},}
\def\IJMPA#1,{{\jnfont Int.\ J.\ Phys.\ A }{\bf #1},}

\begin{document}

\title{\ \\[10mm] Revisiting wrong sign Yukawa coupling of type II two-Higgs-doublet model in light of the recent LHC data}

\author{Lei Wang, Hong-Xin Wang, Xiao-Fang Han$^{*}$\footnotetext{*) Corresponding author.
Email address: xfhan@ytu.edu.cn (X.
Han)}}
\affiliation{Department of Physics, Yantai University, Yantai
264005, China}


\begin{abstract}
In light of the recent LHC Higgs data, we examine the parameter space of type II two-Higgs-doublet model in which the 125 GeV Higgs has the wrong sign Yukawa couplings. Combining related theoretical and experimental limits, we find that
the LHC Higgs data exclude most of the parameter space of the wrong sign Yukawa coupling.
For $m_H=$ 600 GeV, the allowed samples are mainly distributed in several corners and narrow bands of
 $m_A<20$ GeV, 30 GeV $<m_A<120$ GeV, 240 GeV $<m_A<300$ GeV, 380 GeV $<m_A< 430$ GeV, and 480 GeV $<m_A< 550$ GeV.
For $m_A=$ 600 GeV, $m_H$ is required to be less than 470 GeV.
The light pseudo-scalar with a mass of 20 GeV is still allowed in case of the wrong sign Yukawa coupling of 125 GeV Higgs.

\end{abstract}
 \pacs{12.60.Fr, 14.80.Ec, 14.80.Bn}

\maketitle

\section{Introduction}
The two-Higgs-doublet model (2HDM) \cite{2hdm} is a popular extension of the SM by introducing another $SU(2)_L$ Higgs doublet, which contains neutral CP-even Higgs bosons $h$ and $H$, neutral pseudoscalar $A$, and charged Higgs $H^{\pm}$.
There are four typical 2HDMs in which the flavor changing neutral
currents at tree level are absent, namely the type-I \cite{i-1,i-2}, the type II \cite{i-1,ii-2}, the lepton-specific, and the flipped models \cite{xy-1,xy-2,xy-3,xy-4}.
In the type II model, the Yukawa couplings of leptons and down-type quarks can be enhanced by a factor $\tan\beta$. Therefore, the flavor observables and the LHC searching for Higgs can give more strict restrictions to the type II model than the other three models.
In the type II 2HDM, the 125 GeV Higgs can have a wrong
sign Yukawa coupling besides a SM-like coupling. Compared with the SM, at least one of the Yukawa couplings of the 125 GeV Higgs has an opposite sign to the couplings of gauge bosons, which is extensively studied in Refs. \cite{ws-1,ws-2,ws-3,ws-5,ws-6,ws-7,ws-8,
ws-9,ws-9-1,ws-9-2,ws-10,ws-11,ws-12,1701.02678,1909.09035,1910.06269}.

At the beginning of 2017, we used the LHC Higgs data at that time to explore the parameter space of type II 2HDM, and found that the $H/A\to \tau^+\tau^-$ and $A\to hZ$ modes can give strong restrictions on the parameter space of the wrong sign Yukawa coupling \cite{1701.02678}. Very recently, Refs. \cite{1909.09035,1910.06269} examined the parameter space with degenerate heavy Higgs masses in the framework of this model. In this work, we will re-examine the wrong sign Yukawa coupling in the type II 2HDM, and scan over the parameter space extensively by considering the recent ATLAS and CMS Higgs data.

Our work is organized as follows. In Sec. II we introduce the type II 2HDM briefly. In Sec. III we implement detailed
numerical calculations. In Sec. IV, we display the allowed parameter space by considering the relevant theoretical and
experimental restrictions. In Sec. V, we provide our conclusions.

\section{type II two-Higgs-doublet model}
The scalar potential with a softly broken discrete $Z_2$ symmetry is given by
\cite{2h-poten}
\begin{eqnarray} \label{V2HDM} \mathrm{V} &=& m_{11}^2
(\Phi_1^{\dagger} \Phi_1) + m_{22}^2 (\Phi_2^{\dagger}
\Phi_2) - \left[m_{12}^2 (\Phi_1^{\dagger} \Phi_2 + \rm h.c.)\right]\nonumber \\
&&+ \frac{\lambda_1}{2}  (\Phi_1^{\dagger} \Phi_1)^2 +
\frac{\lambda_2}{2} (\Phi_2^{\dagger} \Phi_2)^2 + \lambda_3
(\Phi_1^{\dagger} \Phi_1)(\Phi_2^{\dagger} \Phi_2) + \lambda_4
(\Phi_1^{\dagger}
\Phi_2)(\Phi_2^{\dagger} \Phi_1) \nonumber \\
&&+ \left[\frac{\lambda_5}{2} (\Phi_1^{\dagger} \Phi_2)^2 + \rm
h.c.\right].
\end{eqnarray}
We focus on the CP-conserving case in which all $\lambda_i$ and
$m_{12}^2$ are real.
The two complex Higgs doublets have the hypercharge $Y = 1$:
\begin{equation}
\Phi_1=\left(\begin{array}{c} \phi_1^+ \\
\frac{1}{\sqrt{2}}\,(v_1+\phi_1^0+ia_1)
\end{array}\right)\,, \ \ \
\Phi_2=\left(\begin{array}{c} \phi_2^+ \\
\frac{1}{\sqrt{2}}\,(v_2+\phi_2^0+ia_2)
\end{array}\right).
\end{equation}
In the above formula, $v_1$ and $v_2$ are the electroweak vacuum expectation values (VEVs) with $v^2 = v^2_1 + v^2_2 = (246~\rm GeV)^2$ and $\tan\beta=v_2 /v_1$. After the spontaneous electroweak
symmetry is broken, we get five physical Higgs particles, two neutral CP-even $h$ and $H$, one neutral pseudoscalar $A$, and a pair of charged scalars $H^{\pm}$.

The Yukawa interactions can be given as
 \bea
- {\cal L} &=&Y_{u2}\,\overline{Q}_L \, \tilde{{ \Phi}}_2 \,u_R
+\,Y_{d1}\,
\overline{Q}_L\,{\Phi}_1 \, d_R\, + \, Y_{\ell 1}\,\overline{L}_L \, {\Phi}_1\,e_R+\, \mbox{h.c.}\,, \eea in which
$Q_L^T=(u_L\,,d_L)$, $L_L^T=(\nu_L\,,l_L)$, and
$\widetilde\Phi_{1,2}=i\tau_2 \Phi_{1,2}^*$. $Y_{u2}$,
$Y_{d1}$ and $Y_{\ell 1}$ are $3 \times 3$ matrices.

The neutral Higgs Yukawa couplings normalized to the SM are as follows.
\bea\label{hffcoupling} &&
y_{h}^{f_i}=\left[\sin(\beta-\alpha)+\cos(\beta-\alpha)\kappa_f\right], \nonumber\\
&&y_{H}^{f_i}=\left[\cos(\beta-\alpha)-\sin(\beta-\alpha)\kappa_f\right], \nonumber\\
&&y_{A}^{f_i}=-i\kappa_f~{\rm (for~u)},~~~~y_{A}^{f_i}=i \kappa_f~{\rm (for~d,~\ell)},\nonumber\\
&&{\rm with}~\kappa_d=\kappa_\ell\equiv-\tan\beta,~~~\kappa_u\equiv 1/\tan\beta.\eea

The Yukawa interactions of the charged Higgs are given as,
\begin{align} \label{eq:Yukawa2}
 \mathcal{L}_Y & = - \frac{\sqrt{2}}{v}\, H^+\, \Big\{\bar{u}_i \left[\kappa_d\,(V_{CKM})_{ij}~ m_{dj} P_R
 - \kappa_u\,m_{ui}~ (V_{CKM})_{ij} ~P_L\right] d_j + \kappa_\ell\,\bar{\nu} m_\ell P_R \ell
 \Big\}+h.c.,
 \end{align}
in which $i,j=1,2,3$.

The neutral Higgs couplings with gauge bosons normalized to the
SM are
\beq
y^{V}_h=\sin(\beta-\alpha),~~~
y^{V}_H=\cos(\beta-\alpha),\label{hvvcoupling}
\eeq
with $V$ denoting $W$ or $Z$.

In type II 2HDM, the SM-like Higgs has not only the SM-like coupling but also the wrong sign Yukawa coupling,
\bea
&&y_h^{f_i}~\times~y^{V}_h > 0~{\rm for~SM-like~coupling},~~~\nonumber\\
&&y_h^{f_i}~\times~y^{V}_h < 0~{\rm for~wrong~sign~Yukawa~coupling}.\label{wrongsign}
\eea
In case of the SM-like coupling, the 125 GeV Higgs couplings are very close to those in the SM , which has an alignment limit. Now we introduce the wrong sign Yukawa coupling.
The absolute values of $y_h^{f_i}$ and $y^{V}_h$ should be close to 1.0 because of the restrictions of 125 GeV Higgs signal data. So we obtain
\begin{align}  &y_h^{f_i}=-1+\epsilon,~~y^{V}_h\simeq 1-0.5\cos^2(\beta-\alpha) ~~{\rm for}~ \sin(\beta-\alpha) >0~{\rm and}~\cos(\beta-\alpha) >0~,\nonumber\\
& y_h^{f_i}=1-\epsilon,~~y^{V}_h\simeq -1+0.5\cos^2(\beta-\alpha) ~~{\rm for}~ \sin(\beta-\alpha)<0~{\rm and}~\cos(\beta-\alpha) >0. \end{align}
Here $\mid\epsilon\mid$ and $\mid\cos(\beta-\alpha)\mid$ are much less than 1.
From Eq. (\ref{hffcoupling}), we can get
\begin{align}\label{wrcp}
&\kappa_f=\frac{-2+\varepsilon+0.5\cos(\beta-\alpha)^2}{\cos(\beta-\alpha)}<<-1 ~{\rm for}~ \sin(\beta-\alpha) >0~{\rm and}~\cos(\beta-\alpha) >0~,\nonumber\\
&\kappa_f=\frac{2-\varepsilon-0.5\cos(\beta-\alpha)^2}{\cos(\beta-\alpha)} >>1 ~{\rm for}~ \sin(\beta-\alpha) <0~{\rm and}~\cos(\beta-\alpha) >0~.
\end{align}
In type II 2HDM, the constraints of $B$-meson and $R_b$ require $\tan\beta$ to be greater than 1, which leads to $\kappa_d < -1$, $\kappa_\ell < -1$, and $0 < \kappa_u < 1$.
Therefore, there is no wrong sign Yukawa coupling for the up-type quark and  may exist wrong sign Yukawa couplings of the down-type quark and lepton for $\sin(\beta-\alpha) > 0$ and $\cos(\beta-\alpha) >0$.
Because of the factor "-2" in the numerator in Eq. (\ref{wrcp}), $\cos(\beta-\alpha)$ and $\tan\beta$ in the wrong sign Yukawa coupling region are greater than those in the SM-like coupling region.

\section{Numerical calculations}
We choose the light CP-even Higgs boson $h$ as the
SM-like Higgs with the mass of $125$ GeV. The branching ratio of $b \to s\gamma$ gives stringent restrictions on the charged Higgs mass of the type II 2HDM, which requires $m_{H^{\pm}} > 570$ GeV \cite{bsr570}.

In the calculation, we take account of the following constraints and observables:

\begin{itemize}
\item[(1)] The electroweak precision data and theoretical constraints.
We use the $\textsf{2HDMC}$ \cite{2hc-1} to consider the theoretical
constraints from the vacuum stability, unitarity and perturbativity, and calculate
the oblique parameters ($S$, $T$, $U$). We take the recent fit results for $S$, $T$, $U$ in Ref. \cite{pdg2018}, 
\beq
S=0.02\pm 0.10, ~~T=0.07\pm 0.12,~~U=0.00 \pm 0.09,
\eeq
with correlation coefficients,
\beq
\rho_{ST} = 0.89, ~~\rho_{SU} = -0.54, ~~\rho_{TU} = -0.83.
\eeq

\item[(2)] The heavy-flavor observables and $R_b$ constraints.
We use $\textsf{SuperIso-3.4}$ \cite{spriso} to
calculate the branching ratio of $B\to X_s\gamma$. $\Delta m_{B_s}$ is calculated following the
formulas of Ref. \cite{deltmq}. Furthermore, we consider the $R_b$ constraints of bottom quarks in $Z$ decays,
which is calculated following the formulas of Refs. \cite{rb1,rb2}. Recently, the $R_b$ observable is also considered in
some works on the 2HDM \cite{rb4,rb5}

\begin{table}
\begin{footnotesize}
\begin{tabular}{| c | c | c | c |}
\hline
\textbf{Channel} & \textbf{Experiment [TeV]} & \textbf{Mass range [GeV]}  &  \textbf{Luminosity [fb$^{-1}$]} \\
\hline
 {$gg/b\bar{b}\to H/A \to \tau^{+}\tau^{-}$} & ATLAS 8~\cite{47Aad:2014vgg} & 90-1000 & 19.5-20.3 \\
{$gg/b\bar{b}\to H/A \to \tau^{+}\tau^{-}$} & CMS 8~\cite{48CMS:2015mca} &  90-1000  &19.7 \\
{$gg/b\bar{b}\to H/A \to \tau^{+}\tau^{-}$} & ATLAS 13~\cite{82vickey} & 200-1200 &13.3 \\
{$gg/b\bar{b}\to H/A \to \tau^{+}\tau^{-}$} & CMS 13~\cite{add-hig-16-037} & 90-3200 &12.9 \\
{$gg\to H/A \to \tau^{+}\tau^{-}$} & CMS 13 \cite{1709.07242}& 200-2250   & 36.1 \\
{$b\bar{b}\to H/A \to \tau^{+}\tau^{-}$} & CMS 13 \cite{1709.07242}& 200-2250   & 36.1 \\
 {$b\bar{b}\to H/A \to \tau^{+}\tau^{-}$} & CMS 8 \cite{1511.03610}& 25-80   & 19.7 \\
\hline
 {$b\bar{b}\to H/A \to \mu^{+}\mu^{-}$} & CMS 8 ~\cite{CMS-HIG-15-009} & 25-60 & 19.7 \\
\hline
 {$pp\to H/A \to \gamma\gamma$} & ATLAS 13 \cite{80lenzi} & 200-2400 & 15.4 \\
{$gg\to H/A \to \gamma\gamma$}& CMS 8+13 \cite{81rovelli}& 500-4000 & 12.9 \\
{$gg\to H/A \to \gamma\gamma$~+~$t\bar{t}H/A~(H/A\to \gamma\gamma)$}& CMS 8  \cite{HIG-17-013-pas}& 80-110 & 19.7 \\
{$gg\to H/A \to \gamma\gamma$~+~$t\bar{t}H/A~(H/A\to \gamma\gamma)$}& CMS 13  \cite{HIG-17-013-pas}& 70-110 & 35.9 \\
{$VV\to H \to \gamma\gamma$~+~$VH~(H\to \gamma\gamma)$}& CMS 8 \cite{HIG-17-013-pas}& 80-110 & 19.7 \\
{$VV\to H \to \gamma\gamma$~+~$VH~(H\to \gamma\gamma)$}& CMS 13 \cite{HIG-17-013-pas}& 70-110 & 35.9 \\
\hline

 {$gg/VV\to H\to W^{+}W^{-}$} & ATLAS 8  \cite{55Aad:2015agg}& 300-1500  &  20.3 \\

{$gg/VV\to H\to W^{+}W^{-}~(\ell\nu\ell\nu)$} & ATLAS 13 \cite{77atlasww13}& 300-3000  &  13.2 \\

{$gg\to H\to W^{+}W^{-}~(\ell\nu qq)$} & ATLAS 13 \cite{78atlasww13lvqq}& 500-3000  &  13.2 \\

{$gg/VV\to H\to W^{+}W^{-}~(\ell\nu qq)$} & ATLAS 13   \cite{1710.07235}& 200-3000  &  36.1 \\
{$gg/VV\to H\to W^{+}W^{-}~(e\nu \mu\nu)$} & ATLAS 13   \cite{1710.01123}& 200-3000  &  36.1 \\
\hline

$gg/VV\to H\to ZZ$ & ATLAS 8 \cite{57Aad:2015kna}& 160-1000 & 20.3 \\

$gg\to H \to ZZ(\ell \ell \nu \nu)$ & ATLAS 13 ~\cite{74koeneke4} & 300-1000  & 13.3 \\
$gg\to H\to ZZ(\nu \nu qq)$ & ATLAS 13 ~\cite{75koeneke5} & 300-3000 & 13.2 \\
$gg/VV\to H\to ZZ(\ell \ell qq)$ & ATLAS 13 ~\cite{75koeneke5} & 300-3000 & 13.2 \\
$gg/VV\to H\to ZZ(\ell\ell\ell\ell)$ & ATLAS 13 ~\cite{76koeneke3} & 200-3000 & 14.8 \\

$gg/VV\to H\to ZZ(\ell\ell\ell\ell+\ell\ell\nu\nu)$ & ATLAS 13 ~\cite{1712.06386} & 200-2000 & 36.1 \\
$gg/VV\to H\to ZZ(\nu\nu qq+\ell\ell qq)$ & ATLAS 13 ~\cite{1708.09638} & 300-5000 & 36.1 \\
\hline

\end{tabular}
\end{footnotesize}
\caption{The upper bounds on the production cross-section times the branching ratio of
$\tau^+\tau^-$, $\mu^+\mu^-$, $\gamma\gamma$, $WW$, and $ZZ$ for
the $H$ and $ A $ searches at 95\%  C.L..}
\label{tabh}
\end{table}

\begin{table}
\begin{footnotesize}
\begin{tabular}{| c | c | c | c |}
\hline
\textbf{Channel} & \textbf{Experiment [TeV]} & \textbf{Mass range [GeV]}  &  \textbf{Luminosity [fb$^{-1}$]} \\
\hline
$gg\to H\to hh \to (\gamma \gamma) (b \bar{b})$ & CMS 8  \cite{64Khachatryan:2016sey} & 250-1100  & 19.7 \\

$gg\to H\to hh \to (b\bar{b}) (b\bar{b})$ & CMS 8 \cite{65Khachatryan:2015yea}&   270-1100   & 17.9 \\

$gg\to H\to hh \to (b\bar{b}) (\tau^{+}\tau^{-})$ & CMS 8  \cite{66Khachatryan:2015tha}&  260-350 & 19.7 \\

$gg \to H\to hh \to b\bar{b}b\bar{b}$ & ATLAS 13 ~\cite{84varol} & 300-3000  &  13.3  \\

$gg \to H\to hh \to b\bar{b}b\bar{b}$ & CMS 13 ~\cite{1710.04960} & 750-3000  &  35.9  \\
$gg \to H\to hh \to (b\bar{b}) (\tau^{+}\tau^{-})$ & CMS 13 ~\cite{1707.02909} & 250-900  &  35.9  \\

$pp \to H\to hh $ & CMS 13 ~\cite{1811.09689} & 250-3000  &  35.9  \\

\hline

$gg\to A\to hZ \to (\tau^{+}\tau^{-}) (\ell \ell)$ & CMS 8  \cite{66Khachatryan:2015tha}& 220-350 & 19.7 \\

$gg\to A\to hZ \to (b\bar{b}) (\ell \ell)$ & CMS 8  \cite{67Khachatryan:2015lba} & 225-600 &19.7 \\

$gg\to A\to hZ\to (\tau^{+}\tau^{-}) Z$ & ATLAS 8  \cite{68Aad:2015wra}&220-1000 & 20.3 \\

 {$gg\to A\to hZ\to (b\bar{b})Z$} & ATLAS 8  \cite{68Aad:2015wra}& 220-1000 & 20.3 \\

{$gg/b\bar{b}\to A\to hZ\to (b\bar{b})Z$}& ATLAS 13  \cite{1712.06518}& 200-2000 & 36.1 \\

{$gg/b\bar{b}\to A\to hZ\to (b\bar{b})Z$}& CMS 13  \cite{1903.00941}& 225-1000 & 35.9 \\
\hline

 {$gg\to h \to AA \to \tau^{+}\tau^{-}\tau^{+}\tau^{-}$} & ATLAS 8 ~\cite{1505.01609} & 4-50 & 20.3 \\
{$pp\to  h \to AA \to \tau^{+}\tau^{-}\tau^{+}\tau^{-}$} & CMS 8 ~\cite{1701.02032} &  5-15  &19.7  \\
{$pp\to  h \to AA \to (\mu^{+}\mu^{-})(b\bar{b})$} & CMS 8 ~\cite{1701.02032} &  25-62.5  &19.7 \\
{$pp\to  h \to AA \to (\mu^{+}\mu^{-})(\tau^{+}\tau^{-})$} & CMS 8 ~\cite{1701.02032} &  15-62.5  &19.7  \\

{$pp\to  h \to AA \to (b\bar{b})(\tau^{+}\tau^{-})$} & CMS 13 ~\cite{1805.10191} &  15-60  &35.9 \\
{$pp\to  h \to AA \to \tau^{+}\tau^{-}\tau^{+}\tau^{-}$} & CMS 13 ~\cite{1907.07235} &  4-15  &35.9 \\
\hline
$gg\to A(H)\to H(A)Z\to (b\bar{b}) (\ell \ell)$ & CMS 8  \cite{160302991} & 40-1000 &19.8 \\

$gg\to A(H)\to H(A)Z\to (\tau^{+}\tau^{-}) (\ell \ell)$ & CMS 8  \cite{160302991}& 20-1000 & 19.8 \\
\hline
\end{tabular}
\end{footnotesize}
\caption{The upper bounds on the production cross-section times the branching ratio for the channels
of Higgs-pair and a Higgs production in association with $Z$ at 95\% C.L..}
\label{tabhh}
\end{table}

\item[(3)] The 125 GeV Higgs signal data. We use the version 2.0 of $\textsf{Lilith}$ \cite{lilith} to perform the calculation of $\chi^2$ for the 125 GeV Higgs signal data combining the LHC run-I and run-II data (up to datasets of 36 fb$^{-1}$).
    We are particularly concerned with the surviving samples for
$\chi^2-\chi^2_{\rm min} \leq 6.18$, in which $\chi^2_{\rm min}$
is the minimum of $\chi^2$. These samples are within the $2\sigma$ range in two-dimensional plane of model parameters.

\item[(4)] The LHC searching for additional Higgs bosons. We use the
$\textsf{HiggsBounds-4.3.1}$ \cite{hb1,hb2} to perform the exclusion
limits from the Higgs searches at LEP at 95\% confidence level.

At the LHC run-I and run-II, the ATLAS and CMS have searched the
additional Higgs via its decaying into various SM modes and
some exotic channels. Because of the destructive interference contributions
to $gg\to A$ production which come from the top-quark loop and the bottom-quark loop in the type II 2HDM, the cross section decreases with the increasing $\tan\beta$, and reaches a minimum value for a moderate $\tan\beta$, which is dominated by the bottom-quark loop for a large enough
value of $\tan\beta$. The cross section of $gg\to H$ production not only depends on $\tan\beta$
and $m_H$, but also $\sin(\beta-\alpha)$. We calculate the cross sections for $A$ and $H$ in the
gluon fusion and $b\bar{b}$-associated production at NNLO in QCD via
$\textsf{SusHi}$ \cite{sushi}. The cross sections of $H$
via vector boson fusion process is derived from the data at LHC
Higgs Cross Section Working Group \cite{higgswg}. We use the
$\textsf{2HDMC}$ to calculate the branching ratios of
various decay channels of $A$ and $H$. In Table \ref{tabh} and Table \ref{tabhh},
we show a complete list of the additional Higgs searches considered in this paper.
When 1$\leq \tan\beta \leq 30$, the heavy charged scalar searches
at LHC cannot impose restrictions on the
model for $m_{H^{\pm}}>500$ GeV \cite{mhp500}. So we do not
include the heavy charged Higgs searches.

For the $A\to hZ$ channel, the CMS collaboration also presented result
of $h \to \tau^+\tau^-$ at the 13 TeV LHC with an integrated luminosity of 35.9 fb$^{-1}$ in Ref. \cite{1910.11634}. However, compared to
the results of Refs. \cite{1712.06518,1903.00941}, the decay width $\Gamma_A/m_A$ corresponding to the bound of Ref. \cite{1910.11634} is
not given clearly. Therefore, we do not include the experimental bound of $A\to hZ\to (\tau^+\tau^-)Z$ channel from Ref. \cite{1910.11634}.
\end{itemize}

\section{Results and discussions}
\subsection{The constraints from the oblique parameters and the 125 GeV Higgs signal data}
\begin{figure}[tb]
 \epsfig{file=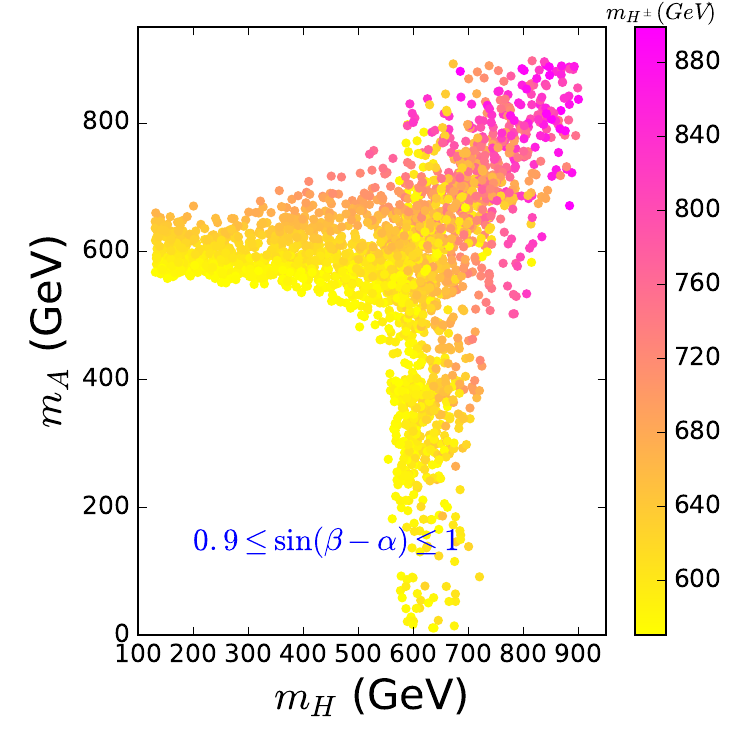,height=7.0cm}
 \epsfig{file=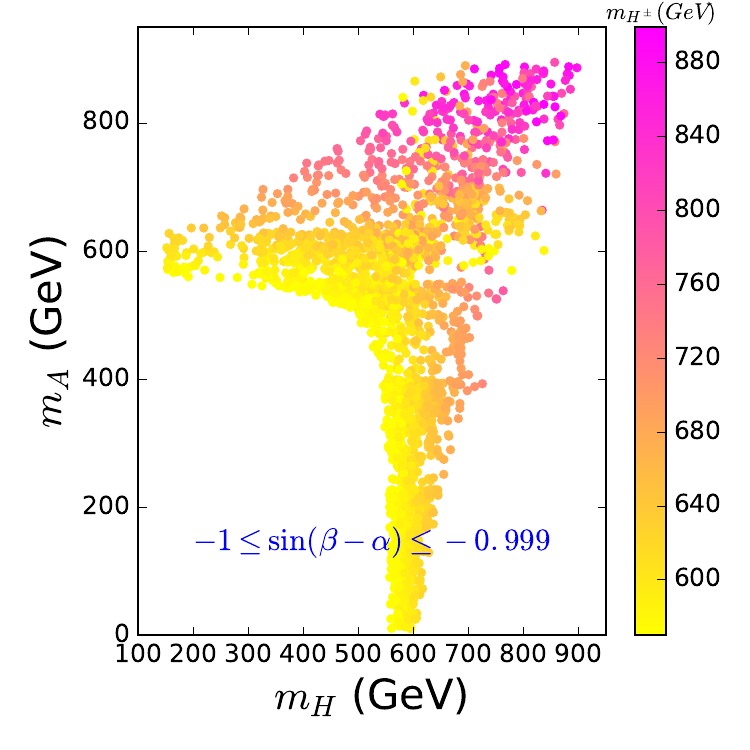,height=7.0cm}
\vspace{-0.0cm} \caption{Scatter plots of $m_A$ and $m_H$ satisfying the constraints of vacuum stability, unitarity, perturbativity, and oblique parameters for 570 GeV $\leq m_{H^{\pm}}\leq$ 900 GeV.} \label{stu}
\end{figure}

In Fig. \ref{stu}, we display the allowed $m_A$ and $m_H$ under the constraints of theory and oblique parameters.
Since the branching fraction of $b \to s\gamma$ imposes a lower bound on the mass of $H^\pm$, $m_{H^{\pm}} > 570$ GeV \cite{bsr570}, we take 570 GeV $\leq m_{H^{\pm}}\leq$ 900 GeV.
When one of $m_A$ and $m_H$ is very closed to $m_{H^{\pm}}$, the contributions of 2HDM to the oblique parameters are sizably suppressed, and the other is allowed to have a large mass splitting with $m_{H^{\pm}}$.
Therefore, as shown in Fig. \ref{stu}, it is unfeasible that
both $m_A$ and $m_H$ are less than 480 GeV, and at least one of $A$ and $H$ is required to have a greater mass.
When one of $m_A$ and $m_H$ is about 600 GeV, the other may have a large mass range, especially for a low mass.
However, when $m_H$ is much greater than 600 GeV and even $m_H=m_{H^{\pm}}$, $m_A$ cannot be very small.
The main reason is from the requirements of vacuum stability,
\beq
\lambda_1>0,~~\lambda_2>0,~~\lambda_3>-\sqrt{\lambda_1\lambda_2}\,,~~\lambda_3+\lambda_4-\mid\lambda_5\mid>-\sqrt{\lambda_1\lambda_2}\,.
\label{vaceq}
\eeq

To better understand the point, we simply assume a very small $\cos(\beta-\alpha)$, and obtain the following relations \cite{ws-9-2},
\begin{eqnarray}
v^2 \lambda_1 &=&  m_h^2 - \frac{t_\beta\,(m_{12}^2 -m_H^2  s_\beta c_\beta ) }{ c_\beta^2}\,,\nonumber \\
v^2 \lambda_2 &=& m_h^2 - \frac{ (m_{12}^2 -m_H^2  s_\beta c_\beta) }{ t_\beta s_\beta^2 }\,,\nonumber\\
\label{eq:alignment2_lambda}
v^2 \lambda_3 &=&  m_h^2 + 2 m_{H^{\pm}}^2 - 2m_H^2 -  \frac{(m_{12}^2 -m_H^2  s_\beta c_\beta)}{  s_\beta c_\beta }\,,\nonumber\\
v^2 \lambda_4 &=&  m_A^2-  2 m_{H^{\pm}}^2 + m_H^2+  \frac{ (m_{12}^2 -m_H^2  s_\beta c_\beta)}{  s_\beta c_\beta }\,,\nonumber\\
v^2 \lambda_5 &=&  m_H^2 - m_A^2+  \frac{ (m_{12}^2 -m_H^2  s_\beta c_\beta)}{  s_\beta c_\beta } \,,
\end{eqnarray}
with $t_\beta\equiv\tan\beta$, $s_\beta\equiv\sin\beta$, and $c_\beta\equiv\cos\beta$.
The first two requirements in Eq. (\ref{vaceq}) are simultaneously satisfied for $m^2_{12}- m^2_H s_\beta c_\beta$ $\to$ 0, and the last two are
respectively satisfied for
\begin{equation}
m_h^2 + m_{H^{\pm}}^2 - m_H^2 > 0\,, \quad \quad m_h^2 + m_{A}^2 - m_H^2  > 0\,.
\label{vaccum}
\end{equation}
The right relation of Eq. (\ref{vaccum}) implies that $m_A$ could not be very small for a very large $m_H$.
The Eq. (\ref{vaccum}) is obtained in the two limits, $\cos(\beta-\alpha)\to 0$ and  $m^2_{12}- m^2_H s_\beta c_\beta\to 0$.
In this paper, we perform exact numerical calculation on the requirements of vacuum stability.
The bounds of Eq. (\ref{vaccum}) can be appropriately loosened by tunning $\cos(\beta-\alpha)$,
$t_\beta$, and $m_{12}^2$.

\begin{figure}[tb]
 \epsfig{file=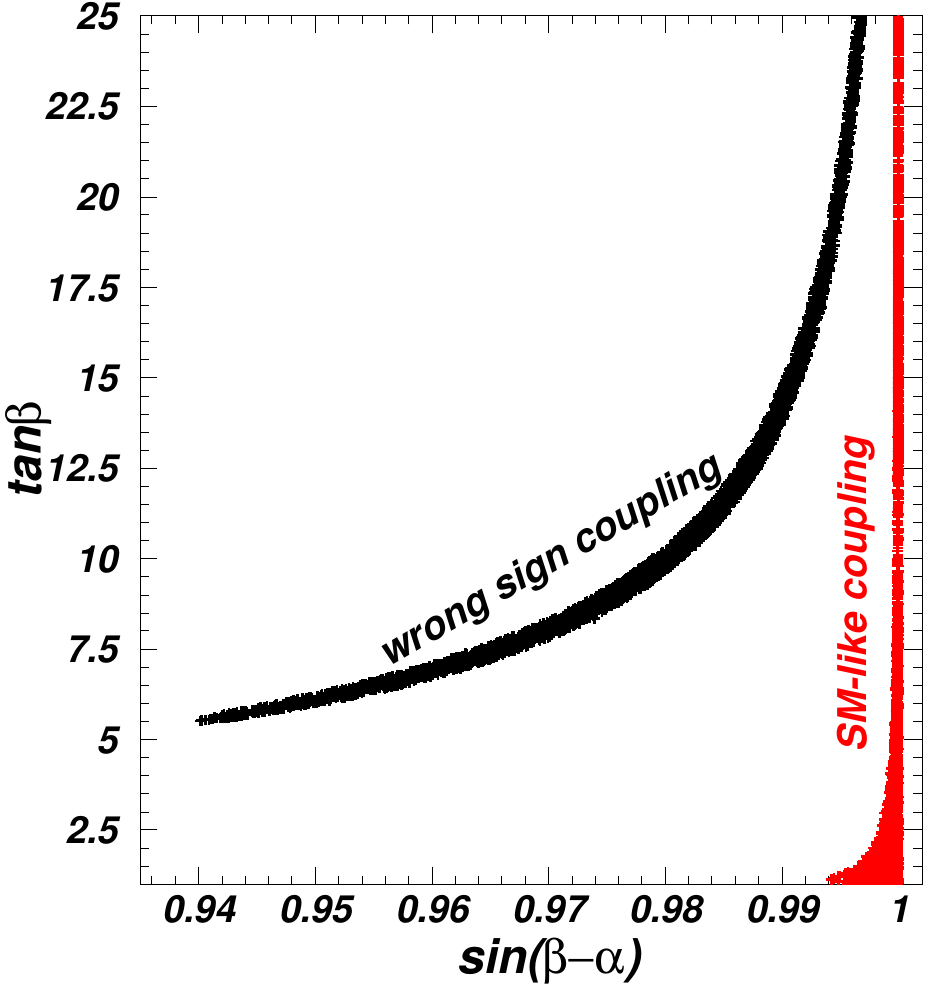,height=7.0cm}
 \epsfig{file=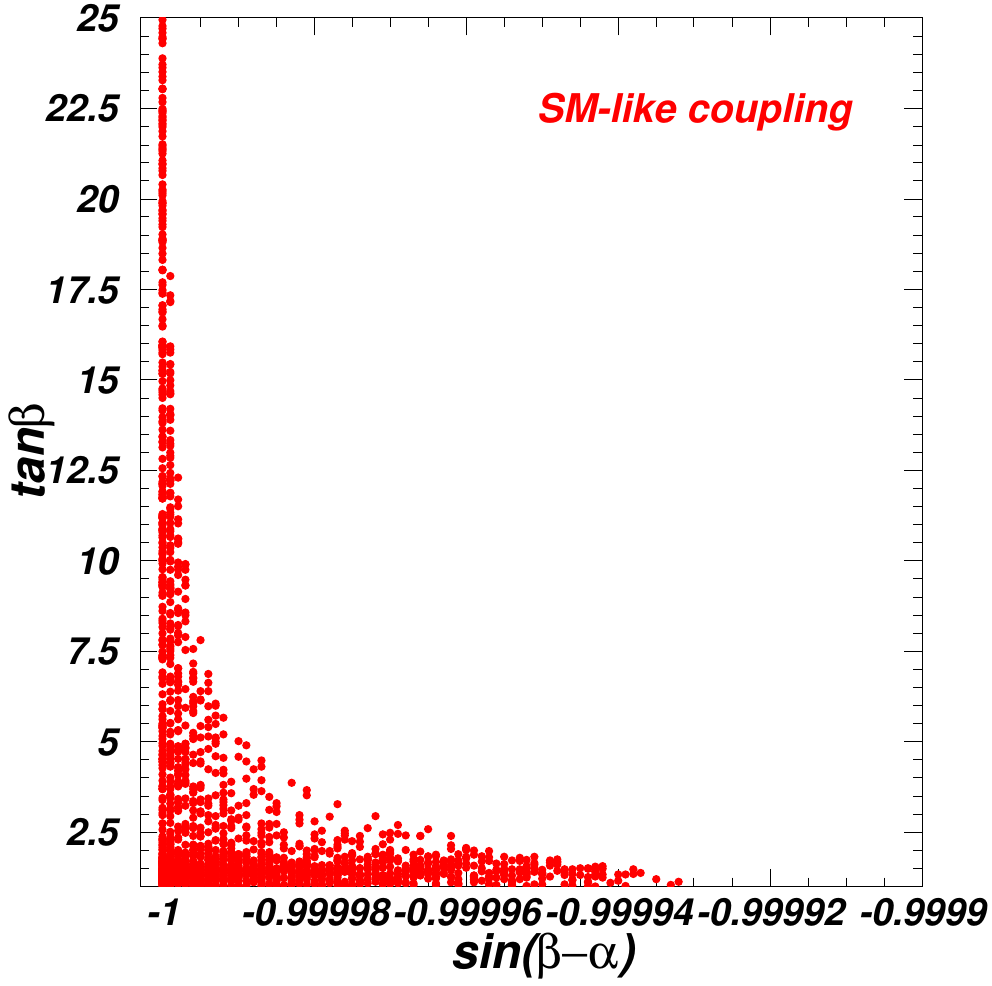,height=7.0cm}
\vspace{-0.0cm} \caption{Scatter plots of $\sin(\beta-\alpha)$ and $\tan\beta$ satisfying the constraints of theory, oblique parameters, and the 125 GeV Higgs signal data.} \label{125higgs}
\end{figure}

Using the survival samples in Fig. \ref{stu} and imposing the restrictions of the 125 GeV Higgs signal data, we obtain the scatter plots of $\tan\beta$ and $\sin(\beta-\alpha)$ in Fig. \ref{125higgs}. From Fig. \ref{125higgs}, we see that the 125 GeV Higgs data can give very stringent constraints on $\tan\beta$ and $\sin(\beta-\alpha)$.
As discussed above, the Yukawa coupling with wrong sign can be achieved only for $\sin(\beta-\alpha) > 0$. In the left panel of Fig. \ref{125higgs}, $\tan\beta$ and $\sin(\beta-\alpha)$ are respectively required to be larger than 5.0 and as low as 0.94 in case of wrong sign couping. When the SM-like coupling is applied, $\sin(\beta-\alpha)$ is restricted to exist in two very narrow bands of $0.994\sim1.0$ and $-1.0\sim-0.99993$, which can be seen in the left and right panels of Fig. \ref{125higgs}.
For a given $\sin(\beta-\alpha)$, $\tan\beta$ is imposed a lower limit in case of the Yukawa coupling with wrong sign, and it is required to be as low as 1.0 in case of the SM-like Higgs coupling.

In order to explicitly show the dependence of $m_A$ ($m_H$) on the other parameters and the specific excluded parameter space from each channel, we do not scan over $m_A$ and $m_H$ simultaneously.
In the following discussions, considering the allowed Higgs mass spectrum shown in Fig. \ref{stu},
we will respectively set $m_A$ or $m_H$ as 600 GeV, and
the other can have a wide mass range, especially for the low mass.
Since heavy Higgs can avoid the restrictions of the LHC direct searches easily, the Higgs with a moderate and low mass
is more interesting.
We scan the parameters for wrong sign Yukawa coupling in the following two scenarios:
 \begin{align}
&0.93\leq \sin(\beta-\alpha) \leq 1.0,~~1\leq \tan\beta \leq 25,~~~570 {\rm\  GeV} \leq ~m_{H^{\pm}}  \leq 900 {\rm\  GeV},\nonumber\\
&{\rm scenario~A:}~~m_H=600~{\rm GeV},~~10~{\rm GeV}\leq m_A\leq~900~ {\rm GeV},\nonumber\\
&{\rm scenario~B:}~~m_A=600~ {\rm GeV},~~150~ {\rm GeV}\leq m_H\leq~900~ {\rm GeV}.
\label{scan}\end{align}
The free parameter $m_{12}^2$ is adjusted to satisfy the theoretical constraint.
Here we take the conventional method \cite{2hc-1}, 0$\leq\beta\leq \frac{\pi}{2}$ and $-\frac{\pi}{2}\leq\beta-\alpha\leq \frac{\pi}{2}$.
Namely, $0 \leq \cos(\beta-\alpha) \leq 1$ and $-1 \leq \sin(\beta-\alpha) \leq 1$.

\subsection{Constraints on scenario A}
\begin{figure}[tb]
 \epsfig{file=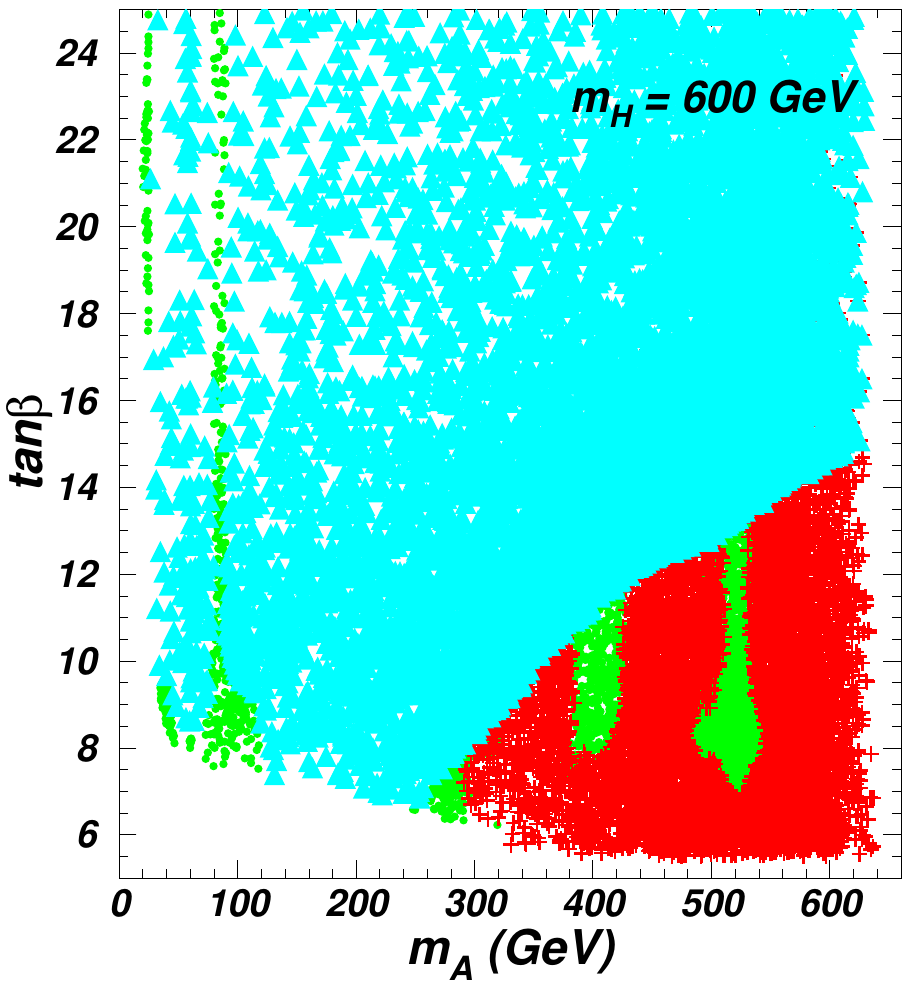,height=7.0cm}
 \epsfig{file=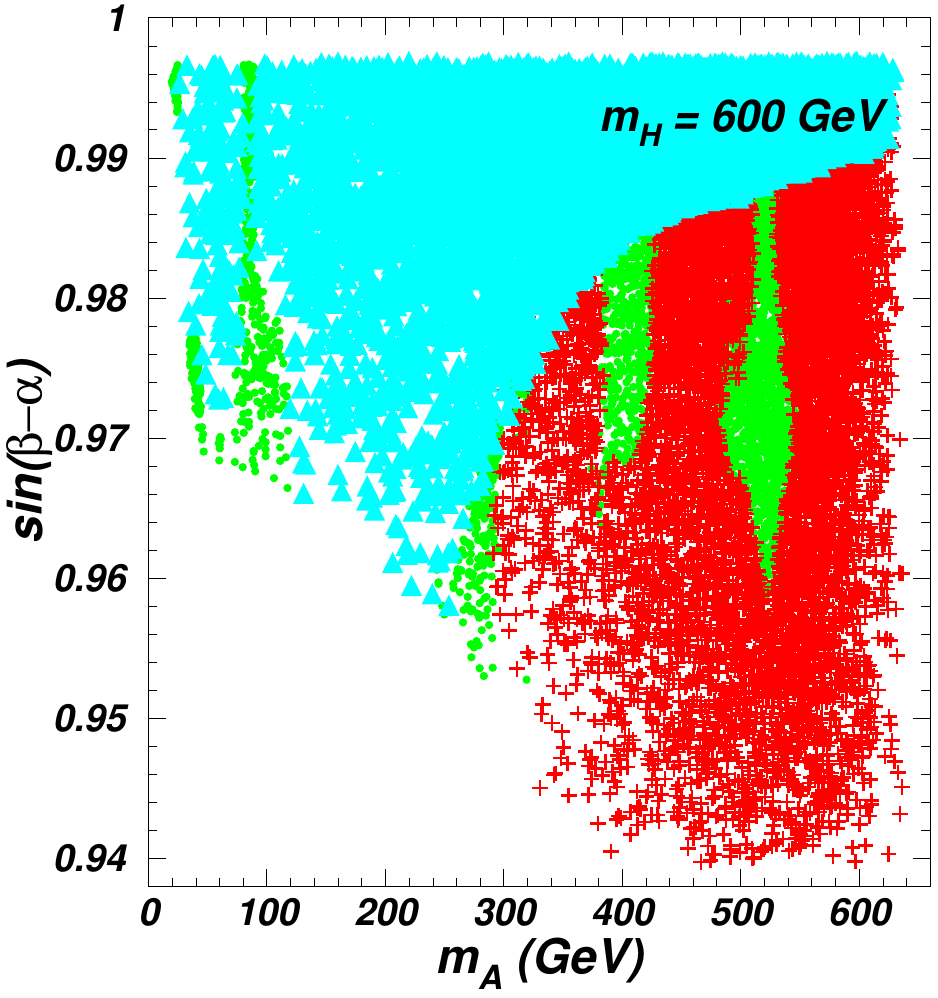,height=7.0cm}
\vspace{-0.3cm} \caption{Scatter plots of $m_A$ versus
$\tan\beta$ and $m_A$ versus $\sin(\beta-\alpha)$ satifying the constraints of pre-LHC and the 125 GeV Higgs signal data. The triangles (sky blue) and pluses (red) are respectively excluded by the $A/H\to \tau^+ \tau^-$ and $A\to hZ$ channels at the LHC.
The bullets (green) are
allowed by various LHC direct searches.} \label{aw}
\end{figure}

Now we extract the allowed parameter space of scenario A after considering the jointly constraints from
pre-LHC (namely the theoretical constraints, electroweak precision
data, the flavor observables, $R_b$, and the exclusions from
searches for Higgs at LEP), the 125 GeV Higgs signal data, and the
searches for additional Higgses at the LHC.
The surviving samples are projected on the planes of $m_A$ versus $\tan\beta$ and $m_A$ versus $\sin(\beta-\alpha)$ in Fig. \ref{aw}.
In case of wrong sign Yukawa coupling, the restrictions mentioned above require $\tan\beta > 5$.
For such range of $\tan\beta$, the cross section of scalar $A$ in the gluon fusion production
is sizably suppressed, and all the samples are favored by the $A\to \gamma\gamma$
and $A\to HZ$ modes.
Since the 125 GeV Higgs signal data give very strict restrictions on the branching ratio of $h\to AA$,
the LHC searching for $h\to AA$ cannot impose constraint on the parameter space.

\begin{figure}[tb]
 \epsfig{file=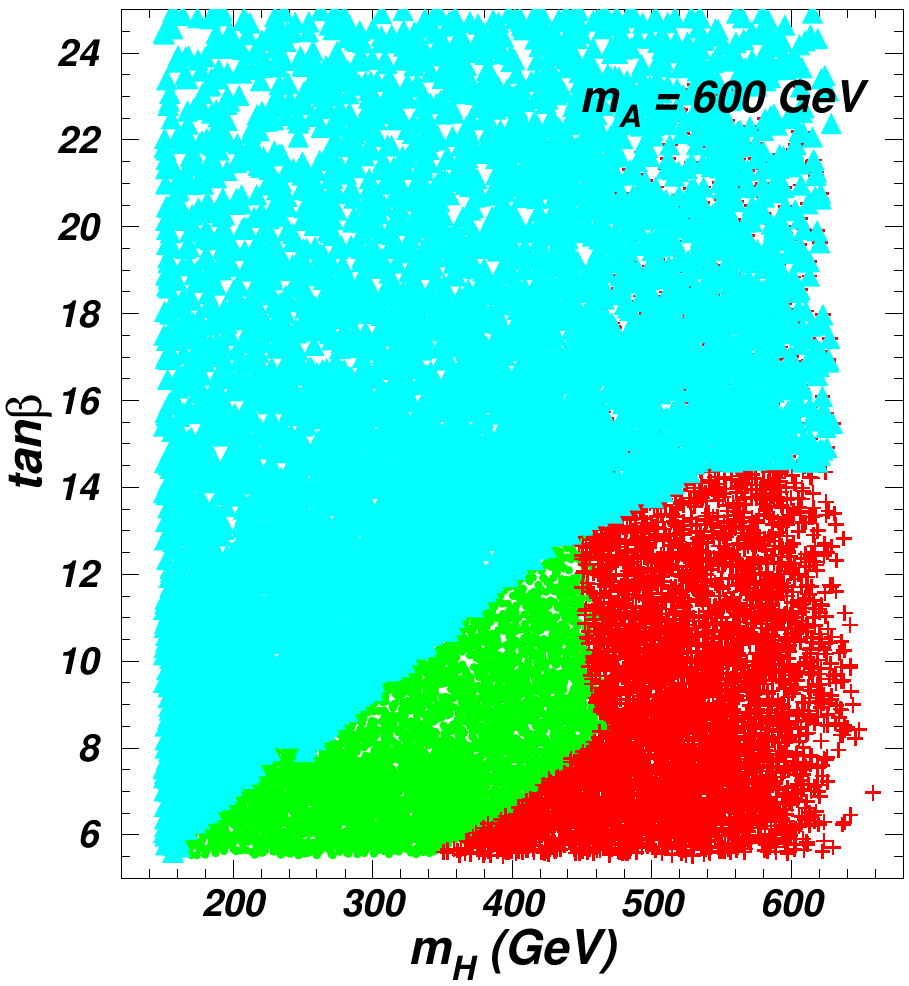,height=7.0cm}
 \epsfig{file=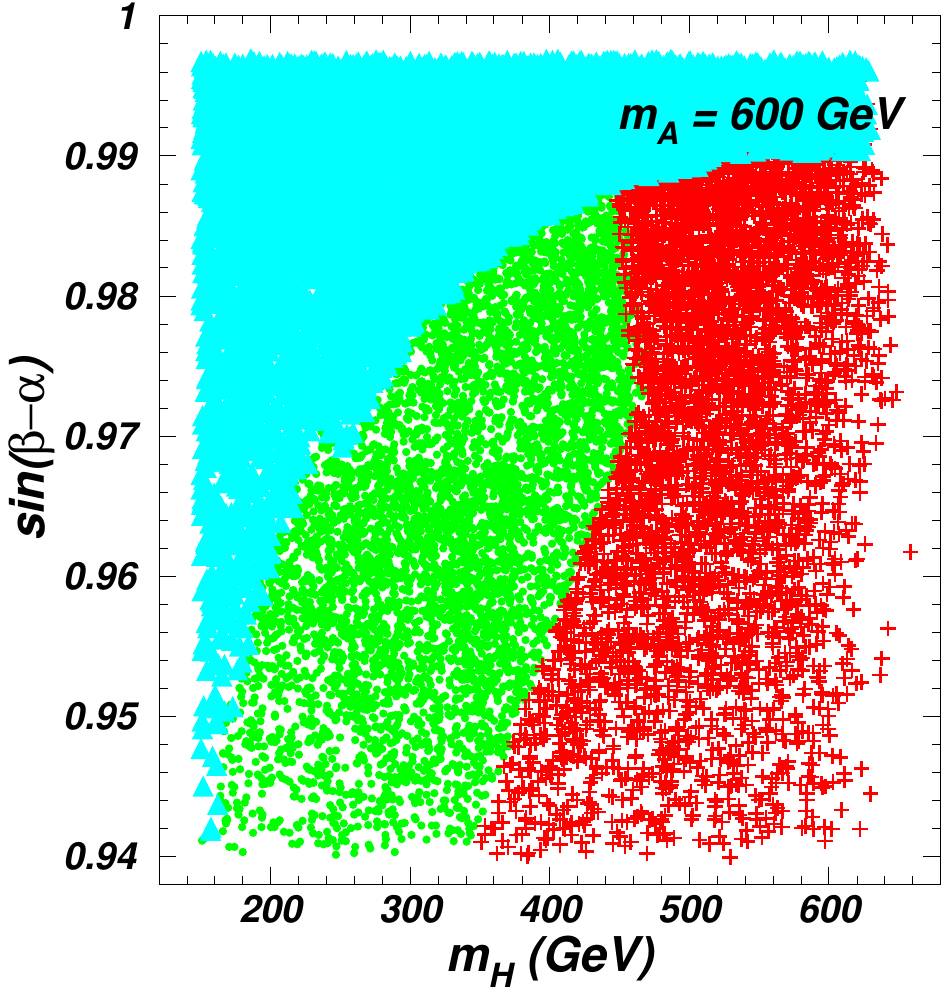,height=7.0cm}
\vspace{-0.3cm} \caption{Scatter plots of $m_H$ versus
$\tan\beta$ and $m_H$ versus $\sin(\beta-\alpha)$ satisfying the constraints of pre-LHC and the 125 GeV Higgs signal data.
The triangles (sky blue) and pluses (red) are
excluded by the $H/A\to \tau^+ \tau^-$ and $A\to hZ$ channels at the LHC respectively.
The bullets (green) are allowed by various LHC direct searches.} \label{bw}
\end{figure}

The $b\bar{b}\to A \to \tau^+ \tau^-$ channel excludes most of the parameter space for large $\tan\beta$ and $gg/b\bar{b}\to A\to hZ$
for small $\tan\beta$. Because the coupling of $AhZ$  is proportional to $\cos(\beta-\alpha)$, the $A\to hZ$ channel tends to exclude
the samples with small $\mid\sin(\beta-\alpha)\mid$.
The allowed samples are mainly distributed in several corners and narrow bands.
As shown in Table \ref{tabh}, the experimental bound of $A \to \tau^+ \tau^-$ channel is absent for
 $m_A<20$ GeV and 80 GeV $<m_A<90$ GeV, and therefore $m_A$ in such mass ranges are allowed. In addition,
 most samples with $m_A$ in the ranges of $30\sim 120$ GeV, $240\sim 300$ GeV, $380\sim 430$ GeV, and $480\sim 550$ GeV
 are allowed for appropriate $\tan\beta$ and $\sin(\beta-\alpha)$. For the last two bands, the experimental bounds of
$A\to hZ$ \cite{1903.00941} are larger than those of neighbouring mass ranges. Therefore,
in the regions of 380 GeV $\leq m_A\leq$ 430 GeV and 480 GeV $\leq m_A\leq$ 550 GeV, many samples with large $\sin(\beta-\alpha)$
can accommodate the bound of $A\to hZ$ channel.

\subsection{Constraints on scenario B}
Here we will study the allowed parameter space in scenario B when imposing the jointly restrictions (1)-(4) in Section III.
The surviving samples can be seen in the scatter plots of $m_H$ versus $\tan\beta$ and $\sin(\beta-\alpha)$ in Fig. \ref{bw}.
Similar to the discussion in scenario A, the pre-LHC and 125 GeV Higgs signal data require $\tan\beta > 5$,
and all samples are favored by the  $H\to VV,~\gamma\gamma,~hh$ and $A\to HZ$ channels.

Fixing $m_A=$ 600 GeV, the channel $b\bar{b}\to H \to \tau^+ \tau^-$ can give upper bounds on $\tan\beta$ and $\sin(\beta-\alpha)$. For instance, $\tan\beta<$ 7.0 (9.2, 14,4) and $\sin(\beta-\alpha)<$ 0.96
(0.98, 0.99) for $m_H=$ 200 GeV (300 GeV, 600 GeV). All samples for $m_H<$ 350 GeV can accommodate the constraints
from the channel $A\to hZ$. For such $m_H$, the mode $A\to HZ$ can increase the total width of $A$, and suppress
the branching ratio of $A\to hZ$ sizably. The channels $A\to\tau^+ \tau^-$ and $A\to hZ$
exclude all the samples for $m_H>$ 470 GeV, and some samples for 150 GeV $<m_H<$ 470 GeV survive for appropriate $\tan\beta$ and $\sin(\beta-\alpha)$.

Compared with the results of Ref. \cite{1701.02678}, the recent LHC Higgs data reduce the parameter space sizably.
For $m_H=600$ GeV, the whole range of $m_A<$ 700 GeV is allowed in Ref. \cite{1701.02678}, while $m_A$ is only allowed to vary in
several ranges in this paper, $m_A<20$ GeV, 30 GeV $<m_A<120$ GeV, 240 GeV $<m_A<300$ GeV, 380 GeV $<m_A< 430$ GeV, and 480 GeV $<m_A< 550$ GeV.
For $m_A=600$ GeV, the whole range of $m_H<$ 700 GeV is allowed in Ref. \cite{1701.02678}, while $m_H<$ 470 GeV is required in the paper. Such differences are mainly caused by the experimental data of $gg/b\bar{b}\to A\to hZ$ from Refs. \cite{1712.06518,1903.00941}, which are not included in Ref. \cite{1701.02678}.

\section{Conclusion}
We have studied the status of wrong sign Yukawa coupling of type II 2HDM in light of recent LHC Higgs data, and obtained some interesting conclusions.
The channels $b\bar{b}\to A/H \to \tau^+ \tau^-$ and $gg/b\bar{b}\to A\to hZ$ exclude most of the parameter space for large $\tan\beta$
and small $\tan\beta$, respectively. For $m_H=$ 600 GeV, the allowed samples are mainly distributed in several corners and narrow bands of
 $m_A<20$ GeV, 30 GeV $<m_A<120$ GeV, 240 GeV $<m_A<300$ GeV, 380 GeV $<m_A< 430$ GeV, and 480 GeV $<m_A< 550$ GeV.
For $m_A=$ 600 GeV, $m_H$ is required to be less than 470 GeV.

\section*{Acknowledgment}
This work was supported by the National Natural Science Foundation
of China under grant 11975013, by the Natural Science Foundation of
Shandong province (ZR2017JL002 and ZR2017MA004).


\begin{thebibliography}{99}
\bibitem{2hdm} T. D. Lee, \PRD8, 1226 (1973).

\bibitem{i-1} H. E. Haber, G. L. Kane and T. Sterling, \NPB161, 493
(1979).

\bibitem{i-2} L. J. Hall and M. B. Wise, \NPB187, 397 (1981).

\bibitem{ii-2} J. F. Donoghue and L. F. Li, \PRD19, 945 (1979).

\bibitem{xy-1} V. D. Barger, J. L. Hewett and R. J. N. Phillips,
\PRD41, 3421 (1990).
\bibitem{xy-2} Y. Grossman, \NPB426, 3 (1994).
\bibitem{xy-3} A. G. Akeroyd and W. J. Stirling, \NPB447, 3 (1995).
\bibitem{xy-4} A. G. Akeroyd, \PLB377, 95 (1996).


\bibitem{ws-1} I. F. Ginzburg, M. Krawczyk, P. Osland, arXiv:hep-ph/0101208.
\bibitem{ws-2} P. M. Ferreira, J. F. Gunion, H. E. Haber and R. Santos, \PRD89, 115003 (2014).
\bibitem{ws-3} B. Dumont, J. F. Gunion, Y. Jiang and S. Kraml, \PRD90, 035021 (2014).
\bibitem{ws-5} D. Fontes, J. C. Romão and J. P. Silva, \PRD90, 015021 (2014).
\bibitem{ws-6} P. M. Ferreira, J. F. Gunion, H. E. Haber, R. Santos, \PRD89, 115003 (2014).
\bibitem{ws-7} D. Fontes, J. C. Romao, J. P. Silva, \PRD90, 015021 (2014).
\bibitem{ws-8} P. M. Ferreira, R. Guedes, M. O. P. Sampaio, R. Santos, \JHEP1412, 067 (2014).
\bibitem{ws-9} L. Wang, X.-F. Han, \JHEP1505, 039 (2015).
\bibitem{ws-9-1} G. C. Dorsch, S. J. Huber, K. Mimasu, J. M. No, Phys. Rev. D 93, 115033 (2016).
\bibitem{ws-9-2} F. Kling, J. M. No, S. Su, JHEP 1609, 093 (2016).

\bibitem{ws-10} A. Biswas, A. Lahiri, \PRD93, 115017 (2016).
\bibitem{ws-11} T. Modak, J. C. Romao, S. Sadhukhan, J. P. Silva, R. Srivastava, \PRD94, 075017 (2016).
\bibitem{ws-12} P. M. Ferreira, S. Liebler, J. Wittbrodt, \PRD97, 055008 (2018).


\bibitem{1701.02678} L. Wang, F. Zhang, X.-F. Han, \PRD95, 115014 (2017).

\bibitem{1909.09035}
W. Su, M. White, A. G. Williams, Y. Wu, arXiv:1909.09035.

\bibitem{1910.06269} W. Su, arXiv:1909.09035.

\bibitem{2h-poten} R. A. Battye, G. D. Brawn, A. Pilaftsis, \JHEP1108, 020 (2011).

\bibitem{bsr570} Heavy Flavor Averaging Group, \EPJC77, 895 (2017);
                 M. Misiak, M. Steinhauser, \EPJC77, 201 (2017).


\bibitem{2hc-1} D. Eriksson, J. Rathsman, O. St{\aa}l, \CPC181, 189 (2010).

\bibitem{pdg2018} M. Tanabashi et al., [Particle Data Group], \PRD98, 030001 (2018).

\bibitem{spriso} F. Mahmoudi, \CPC180, 1579-1673 (2009).

\bibitem{deltmq} C. Q. Geng and J. N. Ng, \PRD38, 2857 (1988)
[Erratum-ibid. D 41, 1715 (1990)].

\bibitem{rb1} H. E. Haber, H. E. Logan, \PRD62, 015011 (2010).

\bibitem{rb2} G. Degrassi, P. Slavich, \PRD81, 075001 (2010).



\bibitem{rb4} N. Chen, J. Gu, T. Han, H. Li, Z. Liu, H. Song, S. Su, W. Su, Y. Wu, J. M. Yang, \IJMPA34, 1940012 (2019).

\bibitem{rb5} N. Chen, T. Han, S. Li, S. Su, W. Su, Y. Wu, arXiv:1912.01431.





\bibitem{lilith} J. Bernon, B. Dumont, S. Kraml, \PRD90, 071301 (2014);
                 S. Kraml, T. Q. Loc, D. T Nhung, L. D. Ninh, arXiv:1908.03952.


\bibitem{hb1} P. Bechtle, O. Brein, S. Heinemeyer, G. Weiglein, K. E.
Williams, \CPC181, 138-167 (2010).

\bibitem{hb2} P. Bechtle, O. Brein, S. Heinemeyer, O. St{\aa}l, T. Stefaniak, G. Weiglein, K. E. Williams,
\EPJC74, 2693 (2014).

\bibitem{sushi} R. V. Harlander, S. Liebler, H. Mantler, \CPC184, 1605 (2013).


\bibitem{higgswg} S. Heinemeyer et al. [LHC Higgs Cross Section Working
Group Collaboration], arXiv:1307.1347.


\bibitem{mhp500} S. Moretti, arXiv:1612.02063.


\bibitem{47Aad:2014vgg}
 ATLAS Collaboration, G.~Aad {\em et~al.}, ``{Search for neutral
  Higgs bosons of the minimal supersymmetric standard model in pp collisions at
  $\sqrt{s}$ = 8 TeV with the ATLAS detector},'' \JHEP11, 056 (2014).

\bibitem{48CMS:2015mca}
 CMS Collaboration,
``{Search for additional neutral Higgs bosons decaying to a pair of tau leptons
  in $pp$ collisions at $\sqrt{s}$ = 7 and 8 TeV},''
CMS-PAS-HIG-14-029.

\bibitem{82vickey}
 ATLAS Collaboration,
``{Search for Minimal Supersymmetric Standard Model Higgs Bosons $H/A$ in the
  $\tau\tau$ final state in up to 13.3 fb$^{-1}$ of pp collisions at
  $\sqrt{s}$= 13 TeV with the ATLAS Detector},''
ATLAS-CONF-2016-085.

\bibitem{add-hig-16-037}
 CMS Collaboration,
``{Search for a neutral MSSM Higgs Boson decaying into $\tau\tau$ $H/A$ with 12.9 fb$^{-1}$ of data at
  $\sqrt{s}$= 13 TeV},''
CMS-PAS-HIG-16-037.

\bibitem{1709.07242}
ATLAS Collaboration,
``{Search for additional heavy neutral Higgs and gauge bosons in the ditau final state produced in 36 fb$^{-1}$
of pp collisions at $\sqrt{s}$= 13 TeV with the ATLAS detector},''
 \JHEP1801, 055 (2018).

\bibitem{1511.03610}
 CMS Collaboration,
``{Search for a low-mass pseudoscalar Higgs boson produced
in association with a $b\bar{b}$ pair in pp collisions at $\sqrt{s}$ = 8 TeV},''
 \PLB758, 296-320 (2016).

\bibitem{CMS-HIG-15-009}
 CMS Collaboration,
``{Search for a light pseudoscalar Higgs boson produced in association with bottom
quarks in pp collisions at $\sqrt{s}$ = 8 TeV},''
CMS-HIG-15-009.


\bibitem{80lenzi}
 ATLAS Collaboration,
``{Search for scalar diphoton resonances with 15.4~fb$^{-1}$ of data collected
  at $\sqrt{s}$=13 TeV in 2015 and 2016 with the ATLAS detector},''
ATLAS-CONF-2016-059.

\bibitem{81rovelli}
 CMS Collaboration,
``{Search for resonant production of high mass photon pairs using
  $12.9\,\mathrm{fb^{-1}}$ of proton-proton collisions at $\sqrt{s} =
  13~\mathrm{TeV}$ and combined interpretation of searches at 8 and 13 TeV},''
CMS-PAS-EXO-16-027.

\bibitem{HIG-17-013-pas}
 CMS Collaboration,
``{Search for new resonances in the diphoton final state in the mass range between
70 and 110 GeV in pp collisions at $\sqrt{s}$ = 8 and 13 TeV},''
CMS-PAS-HIG-17-013.


\bibitem{55Aad:2015agg}
 ATLAS Collaboration, G.~Aad {\em et~al.}, ``{Search for a high-mass
  Higgs boson decaying to a $W$ boson pair in $pp$ collisions at $\sqrt{s} = 8$
  TeV with the ATLAS detector},'' \JHEP01, (2016) 032.


\bibitem{77atlasww13}
 ATLAS collaboration,
``{Search for a high-mass Higgs boson decaying to a pair of W bosons in pp
  collisions at $\sqrt{s}=13$ TeV with the ATLAS detector},'' ATLAS-CONF-2016-074.

\bibitem{78atlasww13lvqq}
 ATLAS Collaboration, ``Search for diboson resonance production
in the $\ell\nu qq$ final state using p p collisions at $\sqrt{s}$ = 13 TeV
with the ATLAS detector at the LHC,''
ATLAS-CONF-2016-062.


\bibitem{1710.07235}
 ATLAS Collaboration,
``Search for WW/WZ resonance production in $\ell\nu qq$ final states in pp collisions
at $\sqrt{s}$ = 13 TeV with the ATLAS detector,''
arXiv:1710.07235.

\bibitem{1710.01123}
ATLAS Collaboration,
``Search for heavy resonances decaying into WW
in the $e\nu\mu\nu$ final state in pp collisions $\sqrt{s}$ = 13 TeV with the ATLAS detector,''
\EPJC78, 24 (2018).



\bibitem{57Aad:2015kna}
 ATLAS Collaboration, G.~Aad {\em et~al.}, ``{Search for an
  additional, heavy Higgs boson in the $H\rightarrow ZZ$ decay channel at
  $\sqrt{s} = 8\;\text{ TeV }$ in $pp$ collision data with the ATLAS
  detector},'' \EPJC76, 45 (2016).

\bibitem{74koeneke4}
 ATLAS Collaboration,
``{Search for new phenomena in the $Z(\rightarrow\ell\ell) +
  E_{\mathrm{T}}^{\mathrm{miss}}$ final state at $\sqrt{s}$ = 13 TeV with thee
  ATLAS detector},''
ATLAS-CONF-2016-056.

\bibitem{75koeneke5}
ATLAS Collaboration,
``{Searches for heavy ZZ and ZW resonances in the $\ell\ell qq$ and vvqq final states in
  pp collisions at $\sqrt{s} = 13$ TeV with the ATLAS detector},''
ATLAS-CONF-2016-082.

\bibitem{76koeneke3}
 ATLAS Collaboration,
``{Study of the Higgs boson properties and search for high-mass scalar
  resonances in the $H \rightarrow ZZ^* \rightarrow 4\ell$ decay channel at
  $\sqrt{s}$ = 13 TeV with the ATLAS detector},''
ATLAS-CONF-2016-079.


\bibitem{1712.06386}
 ATLAS Collaboration,
``{Search for heavy ZZ resonances in the $\ell^+\ell^-\ell^+\ell^-$ and $\ell^+\ell^-\nu\nu$ final states
using proton proton collisions at $\sqrt{s}$ = 13 TeV with the ATLAS detector},''
arXiv:1712.06386.

\bibitem{1708.09638}
ATLAS Collaboration,
``{Searches for heavy ZZ and ZW resonances in the $\ell\ell qq$ and $\nu\nu qq$ final states in pp collisions
at $\sqrt{s}$ = 13 TeV with the ATLAS detector},''
arXiv:1708.09638.




\bibitem{64Khachatryan:2016sey}
 CMS Collaboration, V.~Khachatryan {\em et~al.}, ``{Search for two
  Higgs bosons in final states containing two photons and two bottom quarks},''
\PRD94, 052012 (2016).

\bibitem{65Khachatryan:2015yea}
 CMS Collaboration, V.~Khachatryan {\em et~al.}, ``{Search for
  resonant pair production of Higgs bosons decaying to two bottom
  quark--antiquark pairs in proton--proton collisions at 8 TeV},''
 \PLB749, 560-582 (2015).

\bibitem{66Khachatryan:2015tha}
 CMS Collaboration, V.~Khachatryan {\em et~al.}, ``{Searches for a
  heavy scalar boson H decaying to a pair of 125 GeV Higgs bosons hh or for a
  heavy pseudoscalar boson A decaying to Zh, in the final states with $h \to
  \tau \tau$},'' \PLB755, 217-244 (2016).


\bibitem{84varol}
 ATLAS Collaboration,
``{Search for pair production of Higgs bosons in the $b\bar{b}b\bar{b}$ final
  state using proton$-$proton collisions at $\sqrt{s} = 13$ TeV with the ATLAS
  detector},'' ATLAS-CONF-2016-049.


\bibitem{1710.04960}
 CMS Collaboration,
``{Search for a massive resonance decaying to a pair of Higgs bosons in the four b quark final state in proton-proton collisions
 at $\sqrt{s} = 13$},''
arXiv:1710.04960.

\bibitem{1707.02909}
 CMS Collaboration,
``{Search for Higgs boson pair production in events with two bottom quarks and two tau leptons in proton-proton collisions
at $\sqrt{s} = 13$},''
arXiv:1707.02909.


\bibitem{1811.09689}
 CMS Collaboration,
``{Combination of searches for Higgs boson pair production in proton-proton collisions at
at $\sqrt{s} = 13$},'' \PRL122, 121803 (2019).

\bibitem{67Khachatryan:2015lba}
 CMS Collaboration, V.~Khachatryan {\em et~al.}, ``{Search for a
  pseudoscalar boson decaying into a $Z$ boson and the 125 GeV Higgs boson in
  $\ell^+\ell^-b\overline{b}$ final states},''
  \PLB748, 221-243 (2015).

\bibitem{68Aad:2015wra}
 ATLAS Collaboration, G.~Aad {\em et~al.}, ``{Search for a CP-odd
  Higgs boson decaying to Zh in pp collisions at $\sqrt{s} = 8$ TeV with the
  ATLAS detector},'' \PLB744, 163-183 (2015).

\bibitem{1712.06518}
 ATLAS Collaboration,
``{Search for heavy resonances decaying into a W or Z boson and a Higgs boson in final states with leptons and b-jets
in 36 $fb^{-1}$ of $\sqrt{s}$ = 13 pp collisions with the ATLAS detector},''
 arXiv:1712.06518.

\bibitem{1903.00941}
 CMS Collaboration,
``{Search for a heavy pseudoscalar boson decaying to a Z and a Higgs boson
at $\sqrt{s}$ = 13 TeV},'' \EPJC79, 564 (2019).


\bibitem{1505.01609}
 ATLAS Collaboration,
``{Search for Higgs bosons decaying to aa in the $\mu\mu\tau\tau$ final state in pp collisions
at $\sqrt{s}$= 8 TeV with the ATLAS experiment},'' \PRD92, 052002 (2015).

\bibitem{1701.02032}
 CMS Collaboration,
``{Search for light bosons in decays of the 125 GeV Higgs boson in proton-proton collisions at $\sqrt{s}$= 8 TeV},''
\JHEP1710, 076 (2017).

\bibitem{1805.10191}
 CMS Collaboration,
``{Search for an exotic decay of the Higgs boson to a pair of light pseudoscalars in the final state with two $b$ quarks and two $\tau$
 leptons in proton-proton collisions at $\sqrt{s}$= 13 TeV},'' \PLB785, 462 (2018).

\bibitem{1907.07235}
 CMS Collaboration,
``{Search for light pseudoscalar boson pairs produced from decays of the 125 GeV Higgs boson in final states with two muons and two nearby tracks in pp collisions at at $\sqrt{s}$= 13 TeV},'' arXiv:1907.07235.


\bibitem{160302991}
CMS Collaboration, V.~Khachatryan {\em et~al.},
``{Search for neutral resonances decaying into a Z boson and
a pair of b jets or $\tau$ leptons},''
\PLB759, 369-394 (2016).

\bibitem{1910.11634}
CMS Collaboration, A. M. Sirunyan {\em et~al.},
``{Search for a heavy pseudoscalar Higgs boson decaying into
a 125 GeV Higgs boson and a Z boson in final states with
two tau and two light leptons at $\sqrt{s}$ = 13 TeV},''
arXiv:1910.11634.



\end{thebibliography}
\end{document}